\documentclass[pra,superscriptaddress,twocolumn,floatfix]{revtex4-1}
\usepackage{graphicx}
\usepackage{dcolumn}
\usepackage{bbm}
\usepackage{amsmath}
\usepackage{float}
\usepackage[english]{babel}
\usepackage[english]{babel}
\usepackage{dsfont}
\usepackage[normalem]{ulem}
\usepackage{xcolor}

\begin{document}
\title{Effectiveness of the Krotov method in controlling open quantum systems }

\author{Marllos E. F. Fernandes}
\affiliation{Departamento de F\'isica, Universidade Federal de S\~ao Carlos (UFSCar)\\ S\~ao Carlos, SP 13565-905, Brazil}

% \email{marllos@df.ufscar.br}
\author{Felipe F. Fanchini}
\affiliation{Faculty of Sciences, UNESP - S{\~a}o Paulo State University, 17033-360 Bauru, S{\~a}o Paulo, Brazil}
\author{Emanuel F. de Lima}
\affiliation{Departamento de F\'isica, Universidade Federal de S\~ao Carlos (UFSCar)\\ S\~ao Carlos, SP 13565-905, Brazil}

% \email{emanuel@df.ufscar.br}

\author{Leonardo K. Castelano}
\email{lkcastelano@df.ufscar.br}
\affiliation{Departamento de F\'isica, Universidade Federal de S\~ao Carlos (UFSCar)\\ S\~ao Carlos, SP 13565-905, Brazil}
\date{\today} 
\begin{abstract}

We apply the Krotov method for open and closed quantum systems with the objective of finding optimized controls to manipulate qubit/qutrit systems in the presence of the external environment. In the case of unitary optimization, the Krotov method is first applied to a quantum system neglecting its interaction with the environment. The resulting controls from the unitary optimization are then used to drive the system along with the environmental noise. In the case of non-unitary optimization, the Krotov method already takes into account the noise during the optimization process. We consider two distinct computational task: target-state preparation and quantum gate implementation. These tasks are carried out in simple qubit/qutrit systems and also in systems presenting leakage states. For the state-preparation cases, the controls from the non-unitary optimization outperform the controls from the unitary optimization. However, as we show here, this is not always true for the implementation of quantum gates. There are some situations where the unitary optimization performs equally well compared to the non-unitary optimization. We verify that these situations corresponds to either the absence of leakage states or to the effects of dissipation being spread uniformly over the system, including non-computational levels. For such cases, the quantum gate implementation must cover the entire Hilbert space and there is no way to dodge dissipation. On the other hand, if the subspace containing the computational levels and its complement are differently affected by dissipation, the non-unitary optimization becomes effective.

\end{abstract}

%Escrevendo para não esquecer:
%  Two conditions for the unitary optimization of quantum gates to be 
%equivalente in terms of the mean fidelity
%(i) Sufficiently weak decoherence (system dependent, but weak decoherence is aways desirable in quantum computing)
% (ii) No decoherence-free subspace containing all the computational basis (exemplo de 4 níveis do revisor)

% A condição (i) é necessária, no caso da (ii) existe a possibilidade da otimização unitária fornecer o mesmo da não unitária se a transformação ficar restrita ao subespaço.

% a otimização unitária é boa quando não se pode evitar (na média) a decoerência durante a aplicação dan porta lógica (a gente viu que essa situação acontece).

% mudar o tom de modo a dizer que nós encontramos situações onde a otimização unitária é equivalente à não unitária, satisfazendo as condições (i) e (ii).

\maketitle
\section{Introduction}

The ability to control quantum systems in the presence of environmental noise, which can lead to undesirable effects such as decoherence, is crucial for quantum computing \cite{PhysRevA.103.012404,PhysRevApplied.16.054023,Koch_2016,Roloff:2009:1546-1955:1837}. There are many strategies to enhance the control of quantum systems in the presence of noise: decoherence-free subspaces~\cite{PhysRevLett.81.2594,PhysRevLett.95.130501,doi:10.1126/science.290.5491.498}, dynamical decoupling~\cite{PhysRevLett.82.2417,PhysRevLett.106.240501}, noiseless subsystems~\cite{viola}, and spectral engineering~\cite{PhysRevLett.104.040401}.
Although beneficial, these tactics are restricted to small systems and are typically limited to specific types of environment-system interactions.

Numerical optimization becomes a particularly appealing method to deal with complicated systems and general forms of interactions~\cite{Goerz_2014,Ball_2021,PhysRevA.96.033417,Niu2019,Levy_2018}.
In particular, the Krotov method (KM) has been extensively applied to the control of open systems~\cite{PhysRevLett.107.130404,PhysRevA.85.032321,PhysRevA.91.052315}. 
In fact, cooperative effects of driving and dissipation have been demonstrated in the stochastic version of open quantum systems using the KM~\cite{PhysRevLett.107.130404}. Moreover, the investigation of time-nonlocal non-Markovian master equation by means of the KM has shown a high fidelity implementation of a quantum gate for a qubit system depending on the control dissipation correlation and the memory effects related to the environment~\cite{PhysRevA.85.032321}.

Here, we use the KM to perform state preparation from a known initial state as well as quantum gate implementation for both qubits and qutrits systems. We also investigate the effects of leakage states, which are non-computational levels that are present, {\it e.g.}, in superconducting qubits~\cite{Clarke2008}. We assume that the systems follow a Markovian master equation with dephasing and/or amplitude-damping noise, which are standard archetypal of noisy channels~\cite{Fonseca_Romero_2012}. 
Furthermore, we use two alternative ways to numerically obtain the controls: (i) optimization in the presence of noise, which we refer to as non-unitary optimization, and (ii) optimization in the absence of noise, which we refer to as unitary optimization.  For state-preparation, as could be expected, we verify that the non-unitary optimization outperforms the unitary optimization. However, for quantum gate implementation, we notice that the non-unitary optimization does not always surpass the unitary optimization. In cases where all states are taken into account suffering the effects of noise, the non-unitary optimization and the unitary optimization essentially have the same performance. The cases where the non-unitary optimization has a better performance than the unitary optimization are those where there exists leakage and a subset of states that is less affected by the dissipation. For these cases, the non-unitary optimization is more efficient because it is possible to find a pathway to dodge dissipation, which does not happen for the unitary optimization.

% Our results show that optimal gate implementation in many open systems may be obtained through the unitary optimization, which is independent of the decay rate and noise types aside from being less computationally demanding.

%For quantum gates implementation, we verify that the non-unitary optimization outperforms the unitary optimization only in a few cases. The difference in the mean fidelity in these cases is negligible, probably due to numerical calculations. This result is noteworthy because it shows that optimal gate implementation for open systems is obtained through the unitary optmization, which is independent of the decay rate and noise types. 

\section{Quantum control applied to a quantum system with dissipation}

To consider a dissipative dynamics, we use the well-known Markovian master equation,
\begin{equation}
\dfrac{d\rho}{dt}=-\frac{i}{\hbar}[H,\rho]+\frac{1}{2}\sum_j\gamma_j\left(2L_j\rho L_j^\dagger-L_j^\dagger L_j\rho-\rho L_j^\dagger L_j\right),\label{non-unitary}
\end{equation}
where the first term in the right hand side describes the unitary evolution and the second term accounts for the dissipation. $L_j$ are the Lindblad operators and $\gamma_j$ are the corresponding decay rates.
The Hamiltonian $H=H_0+\varepsilon_\gamma(t) H_1$ can be divided into two terms, where the first term is the constant Hamiltonian $H_0$ and the second term is composed by the product of the time-dependent control function $\varepsilon_\gamma(t)$ and the Hamiltonian $H_1$. Since the optimization approach will be employed for each value of $\gamma$, as detailed below, the control function is dependent on the decay rate. 

%\section{Krotov method}
The optimized controls are sought numerically by the Krotov method~\cite{Goerz_2014}. This method has been employed to study protocols related to standard-gate quantum computing with great {success }~\cite{PhysRevLett.89.188301}. The KM is an iterative monotonic approach for finding optimized controls that minimize a certain functional that is dependent on the control functions and the desired outcome. Here, we employ the same functional adopted in Ref.~\cite{Goerz_2014}, which is given by
\begin{equation}
    J_T=1-\sum_{i=0}^{n-1}\frac{w_i}{\rm{Tr}\left[\rho^2_i(0)\right]}\rm{Re}\left\{
    \rm{Tr}\left[O\rho_i(0)O^\dagger\rho_i(T)\right]\right\},\label{functional0}
\end{equation}
where $w_i$ is a weight for each initial state $\rho_i(0)$, $O$ is the desired unitary operation, and $\rho_i(T)$ corresponds to the {\it i-th} initial state evolved up to the final time $T$. Within this formulation, one might consider a collection of $n-$initial states, each with a different weight, which can speed up the convergence of the KM~\cite{Goerz_2014}. To obtain the control equations of the KM through variational calculus, a further constraint must be added,
\begin{equation}
    J=J_T+\int_0^T\frac{\left(\varepsilon_\gamma(t)-\varepsilon^{ref}_\gamma(t)\right)^2}{\lambda S(t)}dt.\label{functional}
\end{equation}
In the above equation, $\lambda$ is a positive constant, $\varepsilon^{ref}_\gamma(t)$ is a reference function, and $S(t)$ is the envelope function. The extra constraint in Eq.~(\ref{functional}) minimizes the fluence, which represents the integrated power transmitted from the control pulse to the system and its environment. Starting with a trial function $\varepsilon_\gamma^{0}(t)$, we need to solve a set of coupled differential equations to find the correction for the control field $\Delta \varepsilon_\gamma(t)$. The first differential equation is related to the backward evolution (from the final time $t=T$ to the initial time $t=0$) of the co-states $\phi_i(t)$ through the following equation 
\begin{equation}\label{oper}
\frac{\partial \phi^{k}_i(t)}{\partial t}=-\frac{i}{\hbar}[H^{k},\phi^{k}_i]-D^\dagger{[\phi^{k}_i]},
\end{equation}
where
\begin{equation}
D^\dagger{[\xi]}=\frac{1}{2}\sum_j\gamma_j\left(2L^\dagger_j\xi L_j-L_j^\dagger L_j\xi-\xi L_j^\dagger L_j\right),\label{dissipator}
\end{equation}
and the subscript index $i$ is related to the set of initial states that are being optimized, $k$ indicates the {\it k-th} { iteration of the algorithm}, while $H^{k}=H_0+\varepsilon^{k}_\gamma(t)H_1$. Equation~(\ref{oper}) is solved imposing a condition to the co-state at the final time, which is given by $\phi^{k}_i(t=T)=\frac{w_i}{\rm{Tr}\left[\rho^2_i(0)\right]}
    \left[O\rho_i(0)O^\dagger\right]$. Additionally, 
the initial states $\rho_i(0)$ are forward evolved according to the master equation,
\begin{equation}\label{SEqu}
 \dfrac{d\rho_i^{k+1}(t)}{dt}=-\frac{i}{\hbar}[H^{k+1},\rho^{k+1}_i]+D{[\rho^{k+1}_i]},
 \end{equation} 
and the correction of the control function at the {\it k-th} interaction is
\begin{equation}
\varepsilon_\gamma^{k+1}(t)=\varepsilon_\gamma^{k}(t)+ \lambda S(t) \Delta \varepsilon^{k+1}_\gamma(t)\label{fieldn}
\end{equation}
where 
\begin{equation}
\Delta\varepsilon^{k+1}_\gamma(t) =\textrm{Im}\left\{\sum_{i=0}^{n-1}\rm{Tr}\left\{\phi^{k}_i(t)\left[H_1,\rho_i^{k+1}(t)\right] \right\}\right\},\label{fmu}
\end{equation}
 Equations~(\ref{oper}-\ref{fmu}) are solved in a self-consistent way and the value of the functional shown in Eq.~(\ref{functional}) monotonically decreases.

\section{ {Simple Systems}}
We begin our investigation of the optimization of open quantum systems by considering one qubit and one qutrit subjected either to the noise of dephasing or amplitude-damping. The qubit Hamiltonian is $H=H_0+\varepsilon_\gamma(t)H_1$, where $H_0=-\hbar\omega_0\sigma_z$ and $
H_1=\hbar\omega_0\sigma_x$. The Pauli spin matrices in the z- and x-direction are respectively denoted by $\sigma_z$ and $\sigma_x$. The terms of the Hamiltonian describing the qutrit are \begin{equation}\label{1.5}
H_0=-\hbar\omega_0\begin{pmatrix}
1 & 0 & 0\\
0 & 0 & 0\\
0 & 0 & -1
\end{pmatrix},
\end{equation}
and
\begin{equation}\label{2}
H_1=\hbar\omega_0\begin{pmatrix}
0 & 0 & 1\\
0 & 0 & 1\\
1 & 1 & 0
\end{pmatrix}.
\end{equation}
Physically, a Hamiltonian with this kind of structure can be found when three electrons are confined in double quantum dots~\cite{Shi2014,carlos} or in trapped ions~\cite{qutrit_gates}.
To numerically solve Eqs.~(4-8), we first need to define some parameters. We adopt the time scale $\tau=\omega_0^{-1}$ and the final evolution time $T=10\tau$. Also, the initial guess for the control function is set up as $\varepsilon_\gamma(t)=A_0S(t)$, where $A_0$ is the initial amplitude of the trial function and $S(t)$ is the envelope function that smoothly switches on and off the control function, given by
\begin{equation}
S(t)=
\begin{cases}
    \sin^2(\frac{\pi t}{2t_r}),& \text{if } t\leq t_r\\
    1,              & \text{if } t_r < t < T-t_r\\
    \sin^2(\frac{\pi(t-T)}{2t_r}),& \text{if } t\geq T-t_r \\
    
\end{cases}\label{guess}
\end{equation}
 In the above equation, we use $A_0=10^{-2}$ and $t_r=T/30$.

\subsection{State preparation}

First, we investigate the optimized control considering the case where an initial pure state described by a density matrix $\rho_0(0)$ is used to prepare a final state $\rho_0(T)=O\rho_0(0)O^\dagger$, where $O$ is some particular quantum gate. For the qubit and the qutrit, we use $\rho_0(0)=|0\rangle\langle 0|$, where $|0\rangle$ is the lowest energy level state. We take the quantum Fourier transform as an example for the target operator for both qubit and qutrit, which is respectively given by $O=(\sigma_x+\sigma_z)/\sqrt{2}$ and
\begin{equation}
    O=\frac{1}{\sqrt{3}}\begin{pmatrix}
1 & 1 & 1\\
1 & e^{2\pi i/3} & e^{4\pi i/3}\\
1 & e^{4\pi i/3} & e^{8\pi i/3}
\end{pmatrix}.
\end{equation} 
We find the optimized control function $\varepsilon_\gamma^{opt}(t)$ for each decay rate $\gamma$. Here, we have a single $\gamma$ in Eq.~(1) because we start our investigation by considering individual types of noise with the same decay rate. After obtaining the optimized control function, we perform the calculation of the fidelity, which is given by

\begin{equation}
    F=\langle 0|O^\dagger\rho(T)O|0\rangle,\label{fid}
\end{equation} 
where $\rho(T)$ is the solution of Eq.~(\ref{non-unitary}) at the final evolution time T. We use two individual different types of Lindblad operators, related to the dephasing channel and the amplitude-damping channel. For the qubit, the sum in Eq.~(\ref{non-unitary}) contains only the term $j=1$ and the Lindblad operator is either $L_1=\sigma_z$ (dephasing) or $L_1=\sigma_-$ (amplitude-damping), where $\sigma_-=(\sigma_x-i\sigma_y)/2$. For the qutrit, the sum in Eq.~(\ref{non-unitary}) contains two terms $j=1,2$ and the Lindblad operators for dephasing are $L_1=s^1_z$ and $L_2=s^2_z$ whereas for amplitude-damping are $L_1=s^1_-$ and $L_2=s^2_-$, where $s^1_z=|1\rangle\langle1|-|0\rangle\langle0|$, $s^2_z=|2\rangle\langle2|-|0\rangle\langle0|$, $s^1_-=|1\rangle\langle0|$, and $s^2_-=|2\rangle\langle0|$.

% \begin{figure}[!ht]
%     \centering
%     \includegraphics{q3s.pdf}
%     \caption{Fidelity evaluated by Eq.~(\ref{fid}) for a qutrit considering dephasing (panel (a)) and amplitude-damping (panel (b)) errors as a function of the decay rate $\gamma$ using the optimized control function obtained from the unitary (blue dotted curve) and non-unitary dynamics (red solid curve).}\label{qutrit_state}
% \end{figure}

In Figure 1, we plot the fidelity for dephasing (panels (a) and (c)) and amplitude-damping (panels (b) and (d)) as a function of the decay rate $\gamma$ for a qubit and a qutrit. The blue dotted curves are obtained through the following steps: (i) find the optimized field $\varepsilon_0^{opt}(t)$ for $\gamma=0$; (ii) plug this optimized field $\varepsilon_0^{opt}(t)$ into Eq.~(\ref{non-unitary}) for each value of $\gamma$ ; (iii) use the evolved density matrix at the final time to evaluate the fidelity of Eq.~(\ref{fid}). The red solid curves evaluated for the optimized fields $\varepsilon_\gamma^{opt}(t)$ is obtained for the corresponding value of $\gamma$, as explained in section II. For a qubit and a qutrit system, the fidelity obtained for the optimized function $\varepsilon_\gamma^{opt}(t)$ evaluated for each value of $\gamma$ is higher than the fidelity calculated with the unitary optimal function $\varepsilon_0^{opt}(t)$, as expected. For $\gamma/\omega_0=0.01$, the fidelity for a qubit considering the { control function obtained from the non-unitary optimization} is 3.5$\%$ and 0.26$\%$ higher { than the one obtained with unitary optimization}, respectively for dephasing (panel (a)) and amplitude-damping (panel (b)) errors. For a qutrit, the fidelity is 7$\%$ (dephasing) and 0.35$\%$ (amplitude-damping) higher when the dynamics is calculated with the { control function obtained from the non-unitary optimization} and $\gamma/\omega_0=0.01$. Based on results of Figure 1, we conclude that the non-unitary optimization is { more successful in preparing} a desired state for both qubit and qutrit systems; specially, for the qutrit subjected to dephasing (see panel (c) of Fig.~1).

\begin{figure}[!tb]
    \centering
    \includegraphics[width=8.5cm]{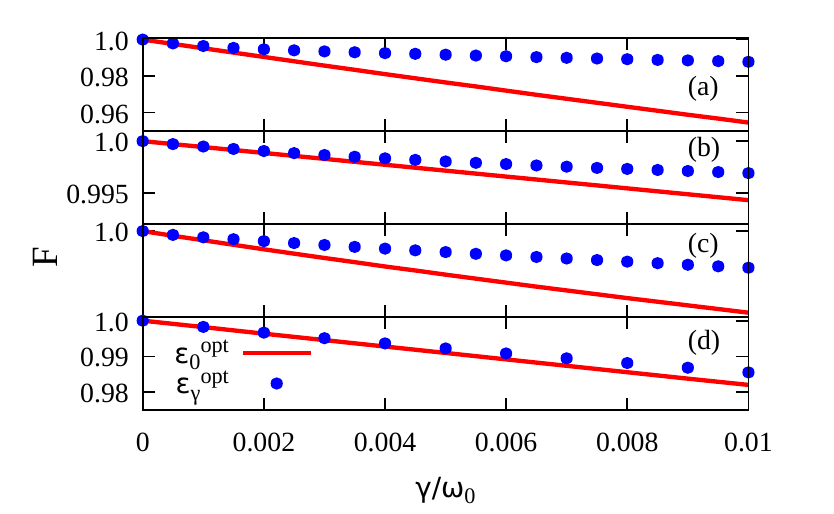}
    \caption{Fidelity evaluated by Eq.~(\ref{fid}) for a qubit considering dephasing (panel (a)) and amplitude-damping (panel (b)) errors as a function of the decay rate $\gamma$ using the optimized control function obtained from the unitary (blue dotted curve) and non-unitary dynamics (red solid curve). The corresponding results for a qutrit are shown in panel (c) (dephasing) and panel (d) (amplitude-damping).}\label{qubit_state}
\end{figure}

These results naturally raise up the question of what would be the physical origin of the observed difference between the optimization carried out with and without the presence of environmental noise. In the unitary optimization, the optimized control for state preparation is obtained within the unitary evolution in the corresponding Hilbert space. With controllability conditions being satisfied, the optimized controls are capable of reaching the highest possible yields. In fact, the solution to the optimal control problem is not unique: there is a myriad of optimized controls leading to the target with corresponding high fidelity yields. Different unitary-optimized controls can lead to distinct values of the final fidelity of the corresponding noisy dynamics, but the unitary optimization cannot distinguish any difference among the optimized controls \textit{a priori}

In the presence of noise, the system non-unitary dynamics occurs in the associated Liouville space (or equivalently in a higher dimensional Hilbert space). Thus, a unitary-obtained control, in general, can no longer steer the system arbitrarily close to the target, since the state space and the dynamics has changed as compared to the unitary case. In the case of non-unitary optimization, the optimization algorithm takes into account the non-unitary structure of the Liouville space, and thus can enhance the performance of the unitary optimization, selecting out the {controls} whose dynamics are least affected by the noise. To support this reasoning, we have numerically verified for the cases and for the range of values of the dissipation considered here that the controls obtained from the non-unitary optimization are optimized controls also for the associated unitary evolution. In the next, we show that this advantage of the non-unitary optimization essentially vanishes for quantum gate implementation when there is no leakage states or when all levels are similarly affected by the noise. 

% \begin{figure}[!ht]
%     \centering
%     \includegraphics{qutrit4s.pdf}
%     \caption{Four states fidelity evaluated by Eq.~(\ref{fidin}) for a qutrit considering dephasing (panel (a)) and amplitude-damping (panel (b)) errors as a function of the decay rate $\gamma$ using the optimized control function obtained from the unitary (green dotted curve) and non-unitary dynamics (black solid curve).}\label{qutrit_4states}
% \end{figure}

%\begin{figure}[!ht]
%    \centering
%    \includegraphics{dephasing_map.eps}
%    \caption{Fidelity as a function of the qubit initial state under dephasing. %(a) unitary optimization (b) non-unitary optimization}\label{damping_map}
%\end{figure}
%\begin{figure}[!ht]
%    \centering
%    \includegraphics{damping_map.eps}
%    \caption{Fidelity as a function of the qubit initial state under amplitude damping. (a) unitary optimization (b) non-unitary optimization}\label{dephasing_map}
%\end{figure}

\subsection{Quantum gate}\label{secB}

We consider the {performance of the optimized control functions} when a quantum gate is the goal of the optimization. This situation is more subtle because the quantum gate should operate over an unknown initial state. To circumvent this situation, the optimization must take into account a set of initial states (for details, see Ref.~\cite{Goerz_2014}).
For qubits, we employ the three initial sates described in Refs.~\cite{regina,palao}, whose matrix elements are given by
\begin{eqnarray}
& & \rho_j(0)=|j\rangle\langle j|\;\textrm{for j=0,1}\nonumber\\
& & \rho_2(0)=\frac{1}{2}\sum_{i,j}|i\rangle\langle j|.\label{iniqubit}
\end{eqnarray} 
For qutrits, we use the following four initial states~\cite{carlos},
\begin{eqnarray}
\rho_j(0)=|j\rangle\langle j|\;\textrm{for j=0,1,2}\nonumber\\
\rho_3(0)=\frac{1}{3}\sum_{i,j}|i\rangle\langle j|.\label{iniqutrit}
\end{eqnarray} 

The weights in Eq.~(\ref{functional0}) are assumed to be $w_j=1/\mathcal{N}$ (unless specified), where $\mathcal{N}=3$ for the qubit and $\mathcal{N}=4$ for the qutrit. {As a first measure of the performance of the control functions, we employ the mean fidelity over {only the} initial states, which is defined as}

\begin{equation}
    F_n=\frac{1}{n}\sum_{i=1}^{n} \left\{
    \rm{Tr}\sqrt{\sqrt{\sigma}\rho_i(T)\sqrt{\sigma}}\right\}^2,\label{fidin}
\end{equation} 
where $\sigma=O\rho_i(0)O^\dagger$, $i$ refers to initial states given by either Eq.~(\ref{iniqubit}) or Eq.~(\ref{iniqutrit}), and $n$ is the corresponding number of initial states.

Figure 2 shows the mean fidelity over initial states for dephasing (panels (a) and (c)) and  amplitude-damping (panels (b) and (d)) as a function of the decay rate $\gamma$, respectively for the qubit and the qutrit. The green dotted curve is obtained for $\varepsilon_0^{opt}(t)$ and evolving the Eq.~(\ref{non-unitary}) with the set of initial states for the qubit Eq.~(\ref{iniqubit}) and for the qutrit Eq.~(\ref{iniqutrit}). The evolved density matrices are used to evaluate the mean fidelity of Eq.~(\ref{fidin}) for each value of $\gamma$. The black solid curves in Figure 2 are evaluated in a similar way, but the employed optimized function $\varepsilon_\gamma^{opt}(t)$ is different for each value of $\gamma$. In Figure 2, one can see that the mean fidelity obtained for the non-unitary optimization surpass the unitary optimization, which numerically shows that the KM is improving the non-unitary control to achieve a higher fidelity for the set of initial states.

\begin{figure}[!ht]
    \centering
    \includegraphics[width=8.5cm]{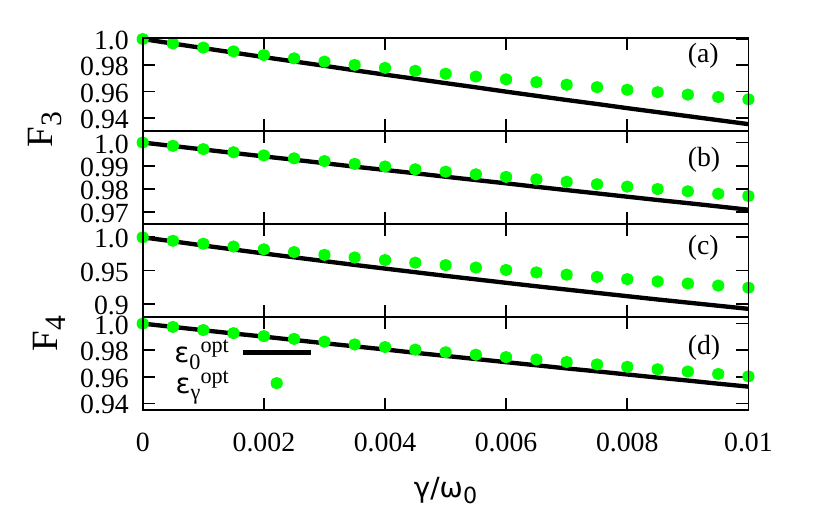}
    \caption{Three states fidelity evaluated by Eq.~(\ref{fidin}) for a qubit considering  dephasing (panel (a)) and amplitude-damping (panel (b)) errors as a function of the decay rate $\gamma$ using the optimized control function obtained from the unitary (green dotted curve) and non-unitary dynamics (black solid curve). The corresponding results for a qutrit, where the four states fidelity is evaluated by Eq.~(\ref{fidin}) considering dephasing and amplitude-damping errors are shown in panels (c) and (d), respectively.}\label{qubit_3states}
\end{figure}

\begin{figure}[!ht]
    \centering
    \includegraphics[width=8.5cm]{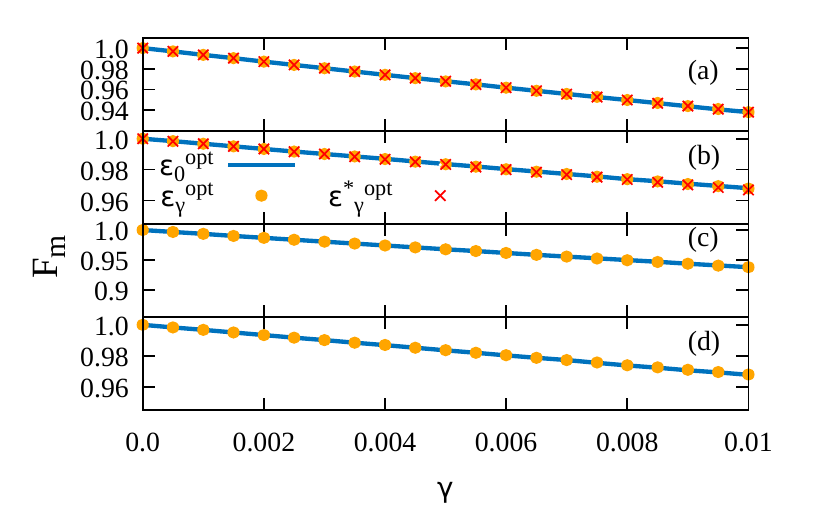}
    \caption{Mean fidelity evaluated by Eq.~(\ref{meanfid}) respectively for a qubit  considering dephasing (panels (a)) and amplitude-damping (panels (b)) errors as a function of the decay rate $\gamma$ using the optimized control function obtained from the unitary (orange dotted curve) and non-unitary dynamics (blue solid curve). The mean fidelity indicated by the red crosses is obtained considering a different set of initial states for only the qubit case, as proposed in Ref.~\cite{Goerz_2014}. Similarly for the qutrit, the mean fidelity is shown in panel (c) (dephasing) and panel (d) (amplitude-damping).}\label{qubit_mean}
\end{figure}

% \begin{figure}
%     \centering
%     \includegraphics{qutrit.pdf}
%     \caption{Mean fidelity evaluated by Eq.~(\ref{meanfid}) for a qutrit considering dephasing (panel (a)) and amplitude-damping (panel (b)) errors as a function of the decay rate $\gamma$ using the optimized control function obtained from the unitary (orange dotted curve) and non-unitary dynamics (blue solid curve).}\label{qutrit_mean}
% \end{figure}

\begin{figure}
    \centering
    \includegraphics[width=8.5cm]{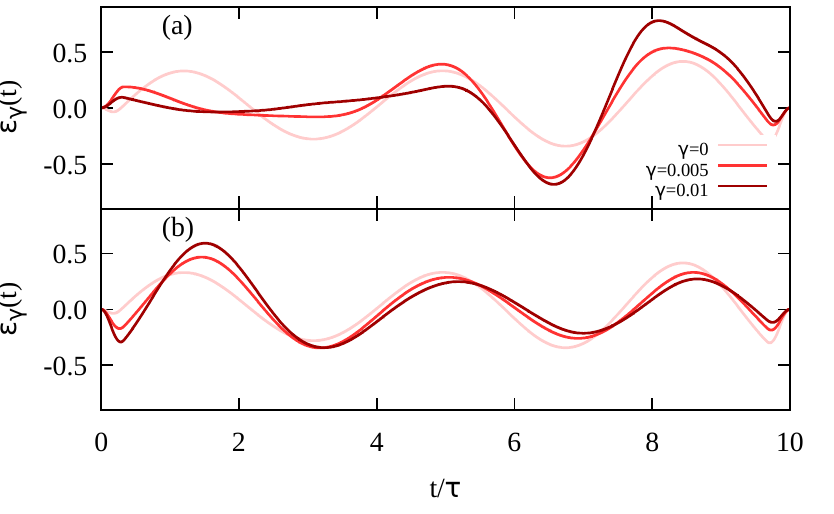}
    \caption{Optimized control functions obtained for a quantum gate implementation for a qubit considering dephasing (panel (a)) and amplitude-damping (panel (b)) for different values of the decay rate $\gamma$.}\label{otimization}
\end{figure}

{ A more pertinent performance criteria for a particular control function in implementing a quantum gate is given by the mean fidelity,} 
\begin{equation}
    F_m=\frac{1}{N_s}\sum_{i=1}^{N_s}\langle \varphi_i|O^\dagger\rho(T)O|\varphi_i\rangle,\label{meanfid}
\end{equation} 
where $|\varphi_i\rangle$ is a pure random state. The ensemble of pure random matrices $|\varphi_i\rangle\langle\varphi_i|$ is built in such a way that all states are uniformly distributed according to the Hilbert-Schmidt norm~\cite{Sommers_2004}.

\begin{figure}[!ht]
    \centering
    \includegraphics[width=8.5cm]{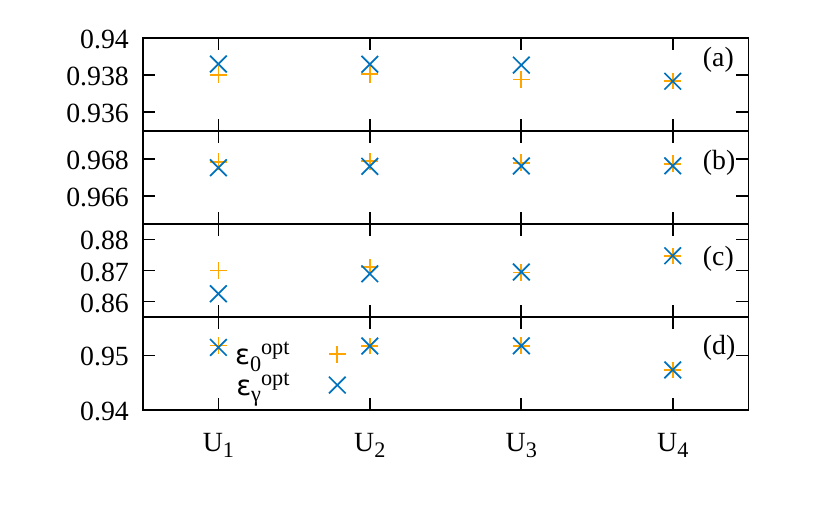}
    \caption{Panels (a) and (c) refer to the dephasing error, while panels (b) and (d) refer to the amplitude-damping error, respectively for the qubti and the qutrit. The results for the mean fidelity for a fixed value of the decay rate $\gamma/\omega_0=0.01$ considering the quantum gates $U_1$, $U_2$, $U_3$, and $U_4$ using the optimized control function obtained from the unitary (orange symbols) and non-unitary dynamics (blue symbols).}\label{qubit_gates}
\end{figure}

% \begin{figure}
%     \centering
%     \includegraphics{qutrit_gates.pdf}
%     \caption{Panels (a) and (b) refer to dephasing and amplitude-damping, respectively. The results for the mean fidelity for a fixed value of the decay rate $\gamma=0.01$ considering one qutrit quantum gates U$_1$, U$_2$, U$_3$ and U$_d$, as defined in the main text. We compare the optimized control function obtained from the unitary (orange symbols) and non-unitary dynamics (blue symbols).}\label{qutrit_gates}
% \end{figure}

Figure 3 shows the mean fidelity for dephasing (panels (a) and (c)) and amplitude-damping (panels (b) and (d)) as a function of the decay rate $\gamma$, respectively for the qubit and the qutrit. The orange dotted curve is obtained for $\varepsilon_0^{opt}(t)$ and evolving the Eq.~(\ref{non-unitary}) considering $N_s=12^4$ initial states. The evolved density matrices are used to evaluate the mean fidelity of Eq.~(\ref{meanfid}) for each value of $\gamma$. The blue solid curves in Figure 3 are evaluated in a similar way, but considering the function $\varepsilon_\gamma^{opt}(t)$, which is optimized for each value of $\gamma$. In Figure 3, one can see that the mean fidelity obtained for the unitary and non-unitary control functions considering both type of errors are essentially the same, although the optimized control functions are distinct for different values of the decay rate (see Fig.~\ref{otimization}).

For the qubit, we also add a result for the mean fidelity considering different initial states proposed in Ref.~\cite{Goerz_2014}, which are given by $\rho_0(0)=2/3|0\rangle\langle0|+1/3|1\rangle\langle1|$,
$\rho_1(0)=1/2(|0\rangle+|1\rangle)(|0\rangle+|1\rangle)$, and $\rho_2(0)=1/2|0\rangle\langle0|+1/2|1\rangle\langle1|$. For this case, we employ the weights in Eq.~(\ref{functional0}) as $w_0=8w_1$ and $w_1=w_2$.
The mean fidelity evaluated for the non-unitary control function $\varepsilon_\gamma^{*opt}(t)$, considering these initial states, is also shown in Figure 3 by the red crosses, but one can observe that the result for the mean fidelity is very similar to the one found considering $\varepsilon_0^{opt}(t)$.

To further investigate this scenario, we consider four different types of quantum gates and evaluate the mean fidelity considering $\gamma/\omega_0=0.01$. Such results are shown in Figure 5. In panels (a) and (b) of Figure 5, we plot the mean fidelity for qubits considering the following quantum gates: $U_1=\sigma_x$, $U_2=\sigma_y$, $U_3=\sigma_z$, and $U_4=\pi/8=|0\rangle\langle0|+e^{i\pi/4}|1\rangle\langle1|$. 

%\textcolor{red}{When the dephasing channel is taking into account (Fig.~5(a)), the mean fidelity evaluated with the non-unitary control is 9\% higher than the one evaluated with the unitary control for the gates $U_1$ and $U_2$. In these cases, the mean fidelity is less than 0.7, which is a range of fidelity where quantum computation cannot be performed.} 
When the dephasing channel is taking into account (Fig.~5(a)), the mean fidelity evaluated with the non-unitary control is higher than the one evaluated with the unitary control for the gates $U_1$, $U_2$, and $U_3$, but the gain is very small, of the order of 0.05\%. On the other hand, the non-unitary optimization provides a {lower} mean fidelity for the amplitude-damping channel (panel (b) of Fig.~5). We believe that this small difference between the unitary and non-unitary optimization seen in Fig.~5 is caused by numerical calculation involved in the KM and in the mean evaluation of Eq.~(\ref{meanfid}), which takes $N_s=12^4$ random initial states.
For qutrits, we use the following quantum gates~\cite{qutrit_gates}:

\begin{equation}
U_1=\frac{1}{\sqrt{2}}
\begin{pmatrix}
1 & -e^{-2\pi i/3} & 0\\
-e^{2\pi i/3} & -1 & 0\\
0 & 0 & -\sqrt{2}
\end{pmatrix},
\end{equation}
\begin{equation}
U_2=\frac{1}{\sqrt{2}}
\begin{pmatrix}
1 & 0 & ie^{2\pi i/3}\\
0 & \sqrt{2} & 0\\
ie^{-2\pi i/3} & 0 & 1
\end{pmatrix},
\end{equation} 
\begin{equation}
U_3=\frac{1}{\sqrt{3}}\begin{pmatrix}
\sqrt{3} & 0 & 0\\
0 & -\sqrt{2} & ie^{-\pi i/6}\\
0 & ie^{\pi i/6} & -\sqrt{2}
\end{pmatrix},
\end{equation}
and \begin{equation}
U_4=\begin{pmatrix}
e^{\pi i/3} & 0 & 0\\
0 & e^{\pi i/6} & 0\\
0 & 0 & e^{-\pi i/2}
\end{pmatrix}.
\end{equation}

To perform these optimizations, we use the same Hamiltonian described in Eq.~(9),  but we have to alter $H_1$ in Eq.~(10) for implementing the gates above for the qutrit. Basically, we used the matrix elements of $H_1$ equal to one in the same position where the quantum gate (Eqs.~(18-21)) has a matrix element different from zero.
In both panels (c) and (d) of Figure 5, we can observe that the non-unitary and the unitary optimizations present an almost identical value for the mean fidelity. 
%\textcolor{red}{In some cases shown in Figure 5, the non-unitary optimization provides a {lower} mean fidelity than the one obtained from the unitary optimization. We believe that this small differences between the unitary and non-unitary optimization seen in Fig. 5 is caused by numerical calculation involved in the KM and in the mean evaluation of Eq.~(\ref{meanfid}), which takes $N_s=12^4$ random initial states.} 
These results show that the unitary optimized control function $\varepsilon_0^{opt}(t)$ is a solution very close to the optimal solution to implement a quantum gate for open quantum systems described by the Markovian master equation in Eq.~(\ref{non-unitary}). We attribute this fact to the constraint of the control field be able to optimize the gate fidelity for all possible initial states. From the evolution of these states, it thus is impossible to avoid the regions of the Hilbert space that are more strongly affected by the noise. Hence, the non-unitary optimization algorithm just searches for a solution to reach the respective set of target states and all other pathways work equally well on average, when compared to the unitary optimization.

\section{Leakage effects}

To further investigate the role of the unitary and non-unitary optimized solutions, we analyze a system that contains non-computational levels, which are also known as leakage states. We compare a two-qubit system, containing only computational states, to a four-levels system that contains only two computational states. We can describe these two systems by the same Hamiltonian, which can be written in the basis of the two-qubit system as
\begin{eqnarray}
    H&=&\hbar\left( J_1\sigma^{(1)}_z+ J_2\sigma^{(2)}_z+  J_{12}\sigma^{(1)}_z\sigma^{(2)}_z\right)\nonumber\\
    &+&\hbar\varepsilon_\gamma(t)(\sigma^{(1)}_x+\sigma^{(2)}_x),
\end{eqnarray}
where $\sigma^{(j)}_m$ is the Pauli spin matrix in the $m$-direction acting on the $j$th-qubit, $J_1$, $J_2$, and $J_{12}$ are parameters describing the time-independent term of $H_0$. The correspondence between states of the two-qubit and four-levels systems is given by: $|0\rangle\leftrightarrow|00\rangle$, $|1\rangle\leftrightarrow|01\rangle$, $|2\rangle\leftrightarrow|10\rangle$, and $|3\rangle\leftrightarrow|11\rangle$.  

The two-qubit system is assumed to be the system free of leakage effects, while the four-levels system presents leakage effects due to action of the quantum gate only in the two lowest energy levels. This difference is clarified by the following arguments. First, we treat a X-gate applied to the first qubit. In the two-qubit system, this gate is described by the operator $O=\sigma_x^{(1)}$, while the X-gate is applied to the two lowest energy levels in the four-levels system, thus $O=|0\rangle\langle1|+|1\rangle\langle0|$. Also, the set of states used in the optimization of the quantum gate is distinct. For the two-qubit system, we must use the set $\{|00\rangle,|01\rangle,|10\rangle,|11\rangle,|\Phi\rangle\}$, $|\Phi\rangle=(|00\rangle+|01\rangle+|10\rangle+|11\rangle)/2$. This set is necessary because the quantum gate act on both qubits, through the X-gate (identity) in the first- (second-) qubit. In the four-level system, the X-gate only acts on the two lowest energy levels, therefore the appropriate set of initial states is similar to the one-qubit system $\{|0\rangle,|1\rangle,|\Phi\rangle\}$, $|\Phi\rangle=(|0\rangle+|1\rangle)/\sqrt{2}$. Another difference between these two systems appears in the mean fidelity calculation. The random initial states for the four-levels system are given by $|\varphi_i\rangle=\alpha_i|0\rangle+\beta_i|1\rangle$, where $|\alpha_i|^2 +|\beta_i|^2=1$ and $\alpha_i$, $\beta_i$ are complex random numbers following the normal distribution. In this case, the quantum gate only acts on the computational subspace relative to the levels $\{|0\rangle,|1\rangle\}$ and there is no constraint imposed on the levels $\{|2\rangle,|3\rangle\}$. For the two-qubit system, the mean fidelity is evaluated using the random initial states $|\varphi_i\rangle=\alpha_i|00\rangle+\beta_i|01\rangle+\gamma_i|01\rangle+\delta_i|01\rangle$, where $|\alpha_i|^2 +|\beta_i|^2+|\gamma_i|^2 +|\delta_i|^2=1$ and $\alpha_i$, $\beta_i$, $\gamma_i$, and $\delta_i$ are complex random numbers sorted according to the normal distribution~\cite{Sommers_2004}. In this case, all states are needed because the quantum gate acts on both qubits simultaneously.

We also probe different types of dissipation to extract more information about the validity of the equivalence of the unitary and non-unitary optimization. The tested cases correspond to the following Lindblad operators that must be plugged into Eq.~(\ref{non-unitary}): (i) $L_1=|0\rangle\langle1|$, $L_2=|0\rangle\langle2|$, and $L_3=|0\rangle\langle3|$, (ii) $L_1=|0\rangle\langle1|$ and $L_2=|0\rangle\langle2|$, (iii) $L_1=|0\rangle\langle2|$ and $L_2=|0\rangle\langle3|$, (iv) $L_1=|0\rangle\langle1|$. When there is more than one Lindblad operator, we use the same decay rate. 
\begin{figure}[!tb]
    \centering
    \includegraphics[width=8.5cm]{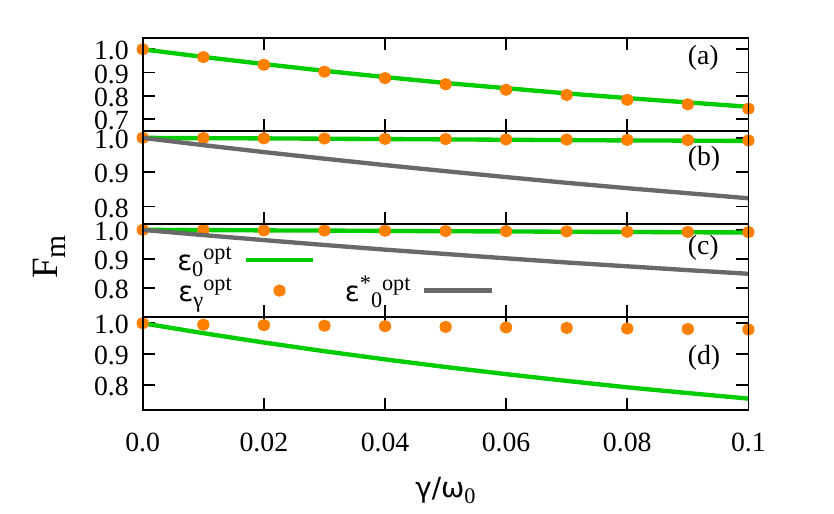}
    \caption{Mean fidelity evaluated by Eq.~(\ref{meanfid}), for four-levels as a function of the decay rate $\gamma$ using the optimized control function obtained from the unitary (green orange curve) and non-unitary dynamics (orange dotted curve). Panels (a) to (d) corresponds to the cases (i) to (iv) of the combinations of the Lindblad operators. The mean fidelity indicated by the dark gray solid line in panels (b) and (c) is obtained considering a different trial control function where the function in Eq.~(\ref{guess}) is multiplied by $\sin{(\omega_{i,0}t)}$, where $\omega_{i,0}$ is the Borh frequency associated to states $|0\rangle$ and $|i\rangle$, where $i=3$ and 2, respectively.}
\end{figure}
\begin{figure}[!tb]
    \centering
    \includegraphics[width=8.5cm]{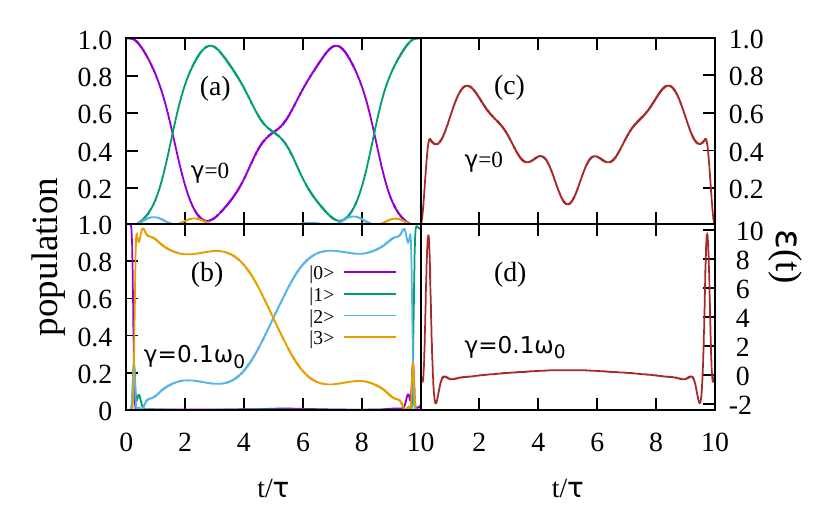}
    \caption{Population of the four-levels system under the dissipation described by the Lindblad operator $L_1=|0\rangle\langle1|$ are shown in panels (a) and (b) as a function of the time for $\gamma=0$ and $\gamma/\omega_0=0.1\omega_0$, respectively. Panels (c) and (d) show the optimized fields for the unitary and non-unitary optimization, respectively.}
\end{figure}
To break the degeneracy between states $|1\rangle$ and $|2\rangle$, we use $J_1/\omega_0=-2$ and $J_2/\omega_0=-0.2$. The coupling between qubits is $J_{12}/\omega_0=0.1$, where $\omega_0=\tau^{-1}$, and the final time is $T=10\tau$. Other parameters not explicitly mentioned here are the same as the ones used in subsection~\ref{secB}.

The cases (i) to (iv) with the above Lindblad operators provide almost the same mean fidelity (results not shown here) considering the unitary and the non-unitary optimization for the two-qubit system. This results are due to the necessity of taking into account all states of the Hilbert space, which does not leave room for dodging dissipation. In panels (a) to (d) of Fig.~6, we plot the mean fidelity for the four-levels system as a function of the decay rate $\gamma$ considering cases (i) to (iv), respectively. The green solid curves show the mean fidelity for the optimized control function obtained from the unitary optimization, while orange dotted curves are relative to the non-unitary optimization. One can see in Fig.~6 that the results for panel (a) corroborate our previous results about the mean fidelity on quantum gates, where the non-unitary and unitary optimizations provide a similar mean fidelity. On the other hand, panels (b) to (d) of Fig.~6 shows that the non-unitary optimization provides a higher mean fidelity than the unitary optimization, when the trial function is properly chosen. For example, in panels (b) and (c) of Fig.~6, the difference between the mean fidelity obtained through the non-unitary and the unitary optimization is rather small when the trial function is proportional to Equation~(\ref{guess}), but it is much more evident when the envelope function of Equation~(\ref{guess}) is multiplied by $\sin{(\omega_{i,0}t)}$, where $\omega_{i,0}$ is the Borh frequency associated to states $|0\rangle$ and $|i\rangle$, where $i=3$ and 2, respectively. These later trial functions are chosen to have the frequency in resonance with the transition between the state $|0\rangle$ and the state $|i\rangle$, which is a state that presents a decay. In other words, when we choose a trial function that induces the population of the dissipative states, the unitary optimization yields a control function that performs the quantum gate with success for $\gamma=0$ without avoiding  dissipative states. On the other hand, the non-unitary optimization yields a control that implements the quantum gate and avoids the dissipation as much as possible to maximize the fidelity no matter the trial function.

The results of Fig.~6 can be better understood from the analysis of Fig.~7, which shows the corresponding population dynamics in panels (a) and (b) evaluated for $\gamma=0$ and $\gamma/\omega_0=0.1$, respectively for the case (iv). The initial state is $|0\rangle$ and the target state is $|1\rangle$. The optimized control field (panel (c)) for $\gamma=0$ drives the system from the initial state to the target state, but it mainly populate the states $|0\rangle$ and $|1\rangle$. When $\gamma/\omega_0=0.1$, the non-unitary optimization search for an optimized control field (panel (d)) that tries to avoid the dissipative state $|1\rangle$, simulated by the Lindblad operator $L_1=|0\rangle\langle1|$, which causes the decay of the population of state $|1\rangle$ to $|0\rangle$. In this situation, the optimization steers the dynamics to a pathway that avoids the state $|1\rangle$ as much as possible.

Other situations that could be probed for the four-level system are the cases where the decay rates between different states are different. These situations can be understood by inspection of the above cases. For instance, cases (ii) to (iv) are limiting cases of case (i), with some decay rates set to zero. Furthermore, if the case (ii) $L_1=|0\rangle\langle1|$ and $L_2=|0\rangle\langle2|$ were considered with different decay rates $\gamma_1$ and $\gamma_2$ in Eq.~(\ref{non-unitary}), we expect a higher mean fidelity for $\gamma_1>\gamma_2$ as compared to the case where $\gamma_1=\gamma_2$ for a fixed value of $\gamma_1$. The limit situation is the one where $\gamma_2=0$, which presents the larger difference between unitary and non-unitary optimization, as can be seen in Fig.~6 (d).

% \begin{figure}[!tb]
%     \centering
%     \includegraphics[width=8.5cm]{fig9new.pdf}
%     \caption{Mean fidelity evaluated by Eq.~(\ref{meanfid}), for four-levels as a function of the decay rate $\gamma$ using the optimized control function obtained from the unitary (green orange curve) and non-unitary dynamics (orange dotted curve). Panels (a) to (d) corresponds to the cases (i) to (iv) of the combinations of the Lindblad operators. The mean fidelity indicated by the dark gray solid line in panels (b) and (c) is obtained considering a different trial control function where the function in Eq.~(\ref{guess}) is multiplied by $\sin{\omega_{i,0}t}$, where $\omega_{i,0}$ is the Borh frequency associated to states $|0\rangle$ and $|i\rangle$, where $i=3$ and 2, respectively.}
% \end{figure}
% To further understand the behavior of the mean fidelity for leaking states, we simulate a situation where the decay rates between different states are different. Thus, we use the scheme (ii) $L_1=|0\rangle\langle1|$ and $L_2=|0\rangle\langle2|$, with different decay rates. In panel (a) of Fig.~(9), we plot the mean fidelity  considering a fixed decay rate $\gamma_f=\gamma_1=0.1\omega_0$ and a varying decay rate $\gamma_v=\gamma_2$ in eq.~(\ref{non-unitary}). In panel (b) of Fig.~(9), we use the opposite configuration between the decay rates, thus $\gamma_f=\gamma_2=0.1\omega_0$ and $\gamma_v=\gamma_1$.

\section{Conclusion}

In this work we used the KM for open and closed quantum systems to numerically investigate state preparation and quantum gate implementation for diverse systems of qubits and qutrits. For the state-preparation, we found that the non-unitary optimization performs better in comparison to the unitary optimization. For quantum gate implementation, on the other hand, unitary and non-unitary optimization results in almost identical mean fidelity when leakage effects are not taken into account. 
%Therefore, we conclude that the non-unitary optimization \textit{does not} perform better than unitary optimization when dealing with the quantum gate implementation.

Our calculations have shown that for state preparation, the result of the non-unitary optimization selects an optimized control from the set of optimized controls of the corresponding unitary dynamics, which performs better in the presence of environmental noise. In other words, the non-unitary optimization chooses a pathway within the dynamics, which is less affected by noise. For the gate implementation, the control function has to operate over all states at the same time since the quantum gate must act over arbitrary input states. Thus, the non-unitary optimization is not anymore able to avoid parts of the Hilbert space that suffers greater influence of the noise. When leakage states are accounted for, the non-unitary optimization is able to present a better performance than the unitary optimization. These fact is related to the existence of a subset of states corresponding to the non-computational subspace that have small decay rates, which opens up a pathway capable of minimizing dissipative effects. Therefore, the inclusion of leakage states with different decay rate plays a fundamental role in the non-unitary optimization, specially when dealing to the quantum gate implementation.

%These situations are interesting because they can provide a scheme to experimentally create systems with levels that are less affected by noise, thereby, helping the quantum gate implementation by using quantum optimal control theory.

%Nevertheless, we conclude that in some relevant situations the best current procedure to implement quantum gates in open quantum systems is to perform unitary optimization. Since this conclusion has been reached from a set of limited cases, additional investigations have to be carried out to verify its extension. Though restricted in some sense, the main finding is relevant since the unitary optimized control functions are independent of decay rates and in general easier to be found numerically than the non-unitary controls. Finally, the present results also indicates that the development of novel techniques for improving non-unitary optimization strategies deserves further exploration.

\begin{acknowledgements}

The authors are grateful for financial support by  the Brazilian Agencies FAPESP, CNPq and CAPES. 
LKC, EFL, and FFF thanks to the Brazilian Agencies FAPESP (grants 2019/09624-3, 2014/23648-9, and 2021/04655-8) and CNPq (grants 311450/2019-9, 423982/2018-4, and 305143/2021-2) for supporting this research. LKC and FFF also acknowledge the National Institute of Science and Technology for Quantum Information (CNPq INCT-IQ 465469/2014-0).
\end{acknowledgements}

\end{document}